\documentclass[prb,twocolumn,superscriptaddress,showpacs,amsmath,amssymb]{revtex4-1}

\usepackage{graphicx}
\usepackage{latexsym}
\usepackage{amsmath}
\usepackage{amssymb}
\usepackage{amsfonts}
\usepackage{color}
\usepackage{bm}
\usepackage{verbatim}
\usepackage{pagecolor,lipsum}

\begin{document}

 \title{ Double-slit Fraunhofer pattern as the signature of the Josephson effect between Berezinskii superconductors through the ferromagnetic vortex. }

\date{\today}

 \author{M.A.~Silaev}
 \affiliation{Department of
Physics and Nanoscience Center, University of Jyv\"askyl\"a, P.O.
Box 35 (YFL), FI-40014 University of Jyv\"askyl\"a, Finland}

 \begin{abstract}
 I apply the recently developed formalism of generalized quasiclassical theory  to show that using 
 hybrid superconducting systems with non-collinear strong ferromagnets one can realize the Josephson junction between Berezinskii-type superconductors.   The reported calculation  reproduces main features observed in the recent experiment, namely the the 
 slightly asymmetric double-slit Fraunhofer interference pattern of the Josephson current through the ferromagnetic vortex. 
 The double-slit structure results from the spatially inhomogeneous Berezinskii state with the amplitude controlled by the 
 local angle between magnetic moments in two ferromagnetic layers. The critical current asymmetry  by the sign of magnetic field 
 can signal the presence of spontaneous supercurrents generated by the non-coplanar magnetic texture near the 
 core of the ferromagnetic vortex core. I demonstrate that ferromagnetic vortex can induce spontaneous vorticity in  
 the odd-frequency order parameter manifesting the possibility  of the emergent magnetic field to create topological defects.
   \end{abstract}

\pacs{} \maketitle

During the recent years large attention has been devoted to the studies of long-range proximity
effect and spin-polarized Josephson currents carried by the spin-triplet Cooper pairs in superconductor/ferromagnet/superconductor (S/F/S) heterostructures\cite{Buzdin2005,Bergeret2005,Eschrig2015a,Linder2015}. 
This interest is motivated the
possible applications of spin-polarized superconducting currents in spintronics. Much effort is invested to the studies of tunable spintronic elements where the Josephson current \cite{Ryazanov2001,Frolov2008,Feofanov2010,Iovan2014,Golubov2002,Robinson2010a,Robinson2010,Anwar2010,Keizer2006} or the
 critical temperature \cite{Tagirov1999,Fominov2003,Zdravkov2013,Fominov2010,Singh2015,Flokstra2010} are controllable by the magnetic degrees of freedom. One of the possible ways for the implementation of such a devise has been suggested in my work \cite{Silaev2009} 
 employing the well-controlled properties of the nanomagnets with vortex-like magnetization patterns. In that proposal I have demonstrated that it is possible to gain the effective control over the long-range proximity effect by tuning the position of ferromagnetic (FM) vortex 
with the help of external in-plane magnetic field. 

 \begin{figure}[h!]
 \centerline{$
 \begin{array}{c}
 \includegraphics[width=0.9\linewidth]{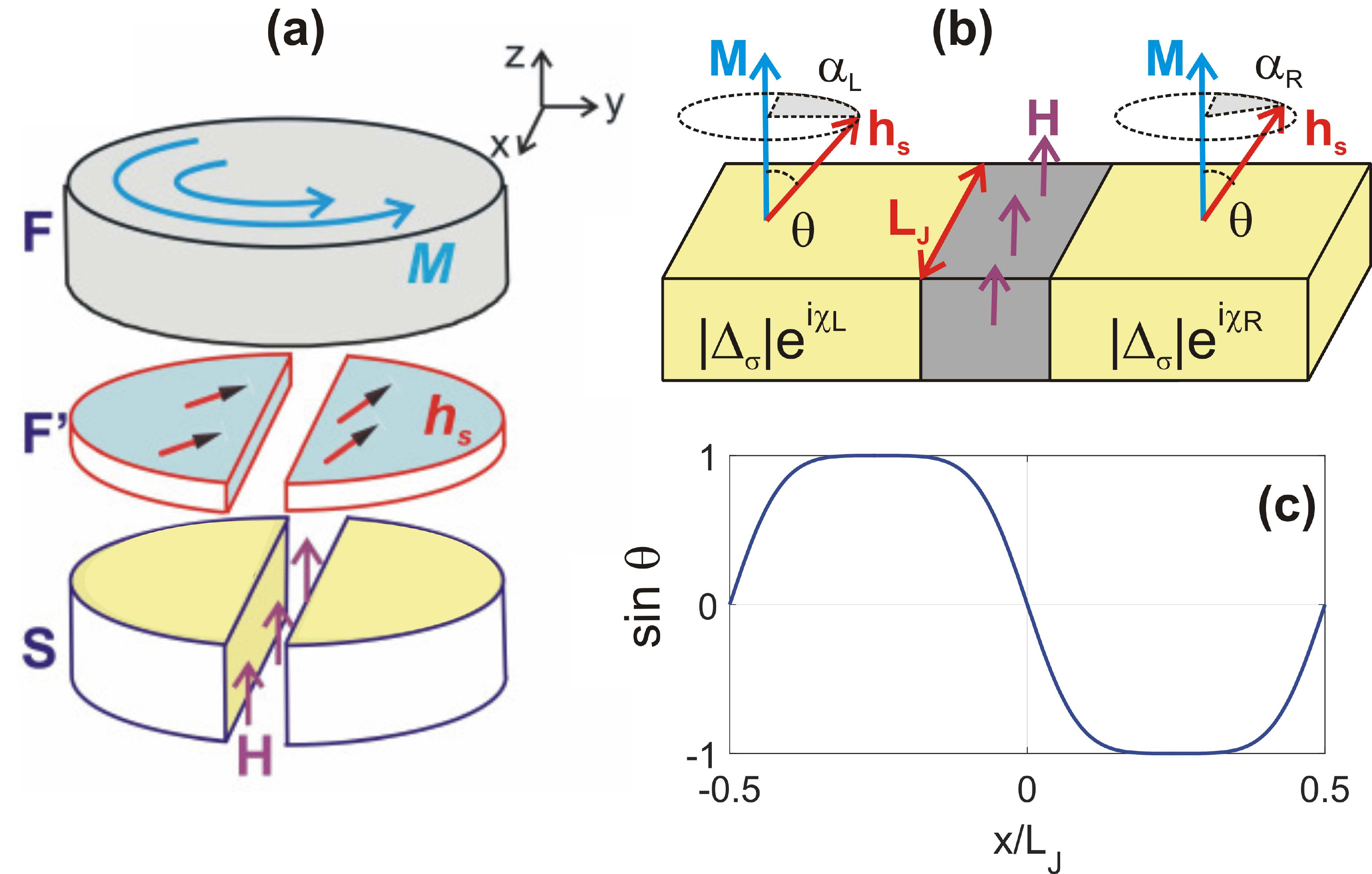} 
 \end{array}$}
 \caption{\label{Fig:SFSmodel} (Color online) 
 (a) Geometry of the Josephson S/F'/F devise with FM vortex in the F layer and the in-plane magnetic texture in the F' layer. 
 (b) Model of the non-homogeneous Josephson junction with usual $\Delta\chi = \chi_R -\chi_L$,
  spin-dependent $\Delta\alpha= \alpha_R - \alpha_L$ phase differences and the texture of the angle 
  $\theta = \theta (x)$ between $\bm M$ and $\bm h_s\parallel \bm M^\prime$. 
  (c) Model dependence of $\sin\theta (x)$ near the junction. }
 \end{figure}

Recently the conceptually similar devise has been realized experimentally\cite{AartsExperiment}. 
Schematically this system is shown in the Fig.(\ref{Fig:SFSmodel})a. It consists of two spin-textured ferromagnets 
F and F' with the magnetizations $\bm M$ and $\bm M^\prime$ and the superconducting layer S. The thin layer F' is in the
contact with superconductor and contains a gap. The Josephson current flows through the layer F which contains FM vortex.
It has been observed that the presence of FM vortex in the system results in the non-trivial modification of the 
critical current dependence $I_c= I_c(H)$ as a function of the external magnetic  field $H$.
The striking features produced by the FM vortex is that this dependence 
becomes similar to the double-slit Fraunhofer interference 
pattern instead of the usual single-slit form produces by the homogeneous junction. 

 I demonstrate below that this behaviour can be considered as
 the direct experimental signatures of the Josephson effect between Berezinskii-type superconductors characterized by the spin-triplet 
 odd-frequency s-wave order parameters 
 \cite{Berezinskii,Kirkpatrick1991,Belitz1992,Belitz1999,Balatsky1992,Abrahams1995,Coleman1994,Fominov2015}. Such superconducting state is 
 induced in the F layer in result of the combined effect of Zeeman splitting in the S electrode and filtering of equal-spin (ES) 
 Cooper pairs in F. Qualitatively the unusual Fraunhofer patterns are explained by the 
 intrinsic inhomogeneity of the induced Berezinskii state which depends on the relative orientation of the 
 local magnetic moments in F and F' layers,  $\bm M$ and $\bm M^\prime$ respectively.  
 The amplitude of induced order parameter is defined by the polar angle $\theta$ between $\bm M$ and $\bm M^\prime$, see Fig.\ref{Fig:SFSmodel}b.
 The variation of $\theta=\theta (x)$ resulting from the magnetic texture along the junction leads to the 
 the experimentally observed  double-slit pattern of the  critical current\citep{AartsExperiment}.
 Besides that, the azimuthal angle $\alpha$ produces spin-dependent phases of the order parameter components.
 The variation of $\alpha$ arises in non-coplanar magnetic textures and in strong ferromagnets with lifted spin degeneracy it produces 
 spontaneous supercurrents \cite{Bobkova2004,Braude2007, Grein2009,Mironov2015, Silaev2017,BobkovaSilaev2017}. 
 The spatial dependence of $\alpha= \alpha (x)$ along the junction is therefore equivalent to the 
 emergent localized magnetic field and results in the critical current asymmetry $I_{c}(H) \neq I_{c}(-H)$ which 
 can signal the presence of spontaneous supercurrents trough the FM vortex.

 The system shown in Fig. \ref{Fig:SFSmodel}a can be described analytically using 
several assumptions. First, the layers are supposed to be thin enough to neglect the variation magnetization fields along the $z$
coordinate. Next, the layer F' is assumed to induce an effective exchange field
within the superconducting electrode $\bm h_s\parallel \bm M^\prime$ . This model is justified by the small thickness of the F'
layer which allows to neglect the variation of correlation functions along z. The Green's functions in F'/S
layer can be found considering just the  superconducting layer with the thickness-averaged exchange field\cite{Bergeret2001}
$\bm h_s = \bm h_{F^\prime}d_{F^\prime}/(d_{F^\prime} +d_S)$ where 
$\bm h_{F^\prime}\parallel \bm M^\prime$ is the exchange field in F', $d_{F^\prime}$ and $d_S$
 are the thickness of the F' and S layers correspondingly. For simplicity the density of states is taken equal in S and F'. 
This exchange field leads to the significant critical temperature suppression of the S/F'/F structure observed in the 
experiment\cite{AartsExperiment}. 
The role of this exchange field is to produce the mixed-spin triplet Cooper pairs which can be converted into the
equal-spin correlations (ESC) in the ferromagnetic layer due to the spin-dependent tunnelling \cite{Silaev2017,BobkovaSilaev2017}.

 The F layer hosting FM vortex has rather large thickness as compared to the mean free path, so that only the the equal-spin correlations (ESC)  
 residing on separate spin-split Fermi surfaces can penetrate to the full thickness. The generalized Usadel equation describing ESC quasiclassical 
 propagators $\hat g_\sigma$ derived in Ref.(\cite{BobkovaSilaev2017}) reads
 \begin{equation} \label{Eq:UsadelGen}
 D_\sigma \hat\partial (\hat g_\sigma \hat\partial \hat g_\sigma) - [\hat\Delta_\sigma+\omega\hat \tau_3, \hat g_\sigma] =0
 \end{equation}
 where $\sigma= \pm 1$ is the spin subband index, $D_\sigma$ are the spin-dependent diffusion coefficients. 
 The covariant differential operator is
  \begin{equation} \label{Eq:DCovariant}
  \hat\partial_{\bm r} = \nabla_{\bm r} + i \sigma {\bm Z}  [\hat\tau_3, \cdot] - 
  i e {\bm A} [\hat\tau_3, \cdot] ,
  \end{equation}
  where $\bm A$ is the electromagnetic vector potential,  
  $\bm Z$ is the adiabatic spin gauge field, and $D_\sigma$ are the spin-dependent diffusion coefficients. 
  The generalized Usadel equation (\ref{Eq:UsadelGen}) is supplemented by the expression for the current
  \begin{equation} \label{Eq:ChargeCurrentDiff}
  {\bm j} = i\pi T e \sum_{\sigma =\pm  } \sum_{\omega }
  \nu_\sigma D_{\sigma}{\rm Tr}[\tau_3 \hat g_{\sigma} \hat\partial_{\bm r}\hat g_{\sigma}] ,
  \end{equation}
  where $\nu_\pm$ is the spin-up/down density of states (DOS).
  
  The effective spin-dependent order parameter $\hat\Delta_\sigma$ in Eq.(\ref{Eq:UsadelGen}) describing the ES components 
  of the proximity-induced order parameter can be obtained from the non-diagonal part of the general tunnelling self-energy  
  \cite{Bergeret2012,Bergeret2012a,Kopnin2011,Kopnin2013}
  \begin{equation} \label{Eq:se}
  \hat\Sigma = \gamma \hat\Gamma \hat F \hat \Gamma^\dagger ,
  \end{equation}
  where $\hat F = (g_{01}\sigma_0 + g_{31} \bm\sigma \bm n_h)e^{i\chi\tau_3}\hat\tau_1$ is the anomalous Green's function in the  
  superconductor with exchange field, $\bm n_h = \bm h_s/h_s$. In the absence of spin-orbital relaxation the spin-singlet and 
  spin-triplet parts are given by 
  $g_{01}= [F_{bcs}(\omega+ih_s)+F_{bcs}(\omega-ih_s)]/2$ and 
  $g_{31} = [F_{bcs}(\omega+ih_s)-F_{bcs}(\omega-ih_s)]/2$, where 
  $F_{bcs}(\omega) = \Delta/\sqrt{\omega^2+ \Delta^2}$.
  The spin-polarized tunnelling matrix has the form 
  $\check \Gamma = t \hat\sigma_0 \hat\tau_0 + u \bm \sigma\bm m \hat\tau_3$ where $\bm m = \bm M/M$ is the direction 
  of magnetization in the ferromagnet and $\gamma = \sigma_n R$ is the parameter describing the barrier strength, 
  $R$ is the normal state tunnelling resistance per unit area. The normalized tunnelling coefficients are
  $t=\sqrt{1+\sqrt{1-P^2}}/2$, $u=\sqrt{1+\sqrt{1-P^2}}/2$ and $P$ is the effective spin-filtering coefficient 
  that ranges from $0$ (no polarization) to $1$ (100 $\%$ filtering efficiency).
  Then the ES component of the tunnelling self-energy is given by
 \begin{equation} \label{Eq:ESse}
 \hat\Sigma_{ES} = \gamma g_{31} e^{i\chi\hat\tau_3}
 [ \hat\tau_1 \bm \sigma \bm h_s + P \hat\tau_2 \bm \sigma (\bm h_s\times \bm m)] .
 \end{equation}
 Projecting $\hat\Sigma_{ES}$ to the spin-up and spin-down states with respect to the quantization axis set 
 by the magnetization $\bm m$ we get the ES components of the effective spin-triplet order parameter
 \begin{equation}\label{Eq:BerezinskyOP}
 \hat \Delta_\sigma = h_\perp \Delta_{F\sigma} \tau_1 e^{i\tau_3 (\chi + \sigma \alpha)} ,
 \end{equation}
 with the amplitude given by $\Delta_{F\sigma} = -( 1+ \sigma P) g_{31}$ and the prefactor
 $h_\perp = \sin \theta$ which is the projection of exchange field to the plane perpendicular to the magnetization direction,
  $\theta$ is the polar angle of $\bm h_s$ in the coordinate system 
 defined by $\bm m$. The distribution of $\sin \theta$ can be obtained by the micromagnetic simulations\cite{AartsExperiment} 
 for the realistic geometry. The additional spin-dependent phase $\alpha$ in Eq.(\ref{Eq:BerezinskyOP}) 
 is defined by the azimuthal angle of $\bm h_s$ as shown in Fig.(\ref{Fig:SFSmodel})b. 
 For the in-plane textures when both $\bm m$ and $\bm h_s$ lie in the $xy$ plane the additional phase in 
 Eq.(\ref{Eq:BerezinskyOP})is absent $\alpha =0 $.  
However near FM vortex core the texture becomes non-coplanar which leads to 
the gradients of $\alpha$, which are coupled to the spin gauge field\cite{Volovik1987,BobkovaSilaev2017} $\bm Z$ so that the combination 
 \begin{equation} \label{Eq:SuperfluidVelocity}
 \bm V_s = \sigma( \nabla \alpha - 2\bm Z)
 \end{equation}
 has the meaning of the texture-induced part of the superfluid velocity. 
In strong ferromagnets with broken spin degeneracy $\bm V_s$ produces spontaneous charge supercurrents\cite{Bobkova2004,BobkovaSilaev2017} 
resulting in the shift of the Fraunhoffer pattern as shown below. 

 Due to the symmetry $g_{31} (\omega) =  - g_{31} (-\omega)$ the pairing amplitude in Eq. (\ref{Eq:BerezinskyOP}) represents 
 the odd-frequency spin-triplet s-wave superconducting  order parameter suggested by Berezinskii \cite{Berezinskii} and intensively studied 
 afterwards\cite{Kirkpatrick1991,Belitz1992,Belitz1999,Balatsky1992,Abrahams1995,Coleman1994,Fominov2015}. 
 The odd superconducting correlations has been studied in  several setups with proximity- induced superconductivity in ferromagnets  \cite{Bergeret2001a,Bergeret2005}, 
 normal metals\cite{Tanaka2007a,Tanaka2007,Tanaka2007b}, topological insulators
 \cite{Yokoyama2012,Black-Schaffer2012} and non-equilibrium systems\cite{Triola2016}.  
 Usually due to the broken translational or/and spin-rotation symmetries the total Cooper wave 
 function in proximity systems is a superposition of odd- and even- frequency components which inevitably coexist at one and the same 
 point\cite{Yokoyama2007,Tanaka2007a,Tanaka2007,Tanaka2007b,Yokoyama2012,Black-Schaffer2012,Triola2016}, except the discrete set of points in the 
 cores of proximity-induced vortices\cite{Alidoust2017}.
 In contrast, the order parameter in Eq.(\ref{Eq:BerezinskyOP}) 
 represents the pure Berezinskii state without admixtures of spin-singlet and/or  even-frequency components. 
 Therefore the setup shown in Fig.(\ref{Fig:SFSmodel})a consisting of two non-collinear strong ferromagnets emulates 
 the Josephson effect between two Berezinskii superconductors through the FM vortex. 
 Note that the absence of even-frequency spin-singlet pairings in the Eq.(\ref{Eq:BerezinskyOP}) is not exact, 
 since the singlet correlations can strictly speaking penetrate even to the strong ferromagnetic layers. 
 However in the dirty regime the singlet amplitude is exponentially suppressed at 
 distances larger than the mean free path from the F/S interface, so that the presence of such components does not affect  
 transport properties in the thick ferromagnetic layer.  
 
 Experimentally the signature of Josephson current between proximity-induced Berezinskii superconductors can be 
 obtained due to the non-trivial structure of the order parameter (\ref{Eq:BerezinskyOP}), which amplitude depends on the 
 angle $\theta$ between magnetic moments in F and F' layers. Such a dependence is peculiar for the spin-triplet 
 odd-frequency pairings since  the spin-singlet component is not sensitive to the exchange field rotations. As shown below, 
 that results in the striking modification of the Fraunhoffer pattern in the critical current as a function of external magnetic field 
 $I_c=I_c(H)$ which reflects the inhomogeneity of the magnetic texture in the S/F'/F system. 
 This behaviour coincides qualitatively with recent experimental observations\cite{AartsExperiment}.
 
 \begin{figure}[h!]
 \centerline{$
 \begin{array}{c}
 \includegraphics[width=1.0\linewidth]{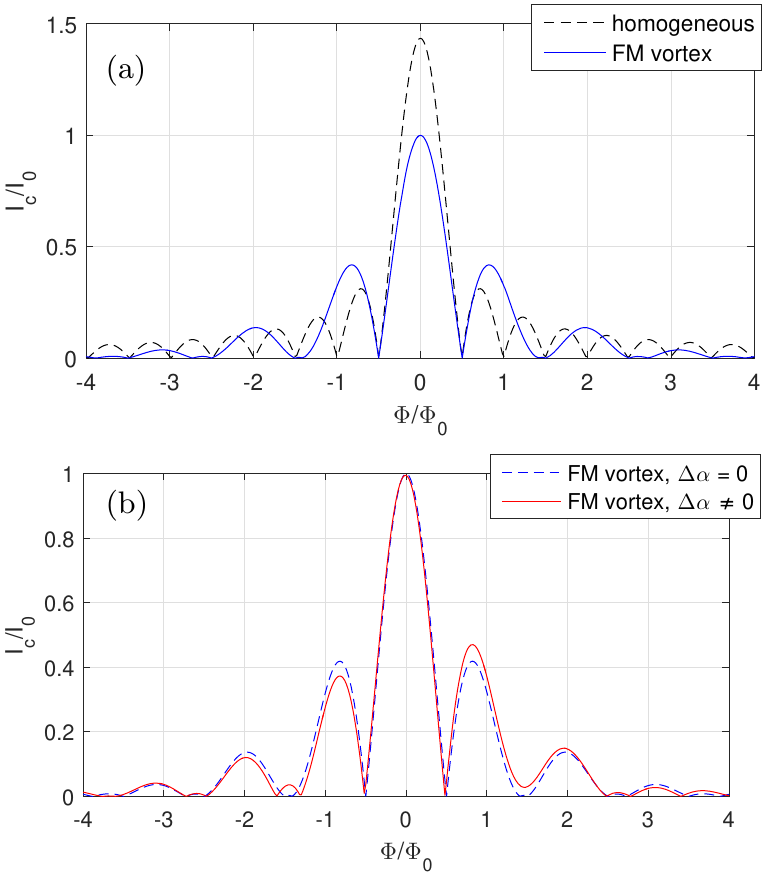}
 \end{array}$}
 \caption{\label{Fig:Jos} (Color online) 
 Interference patterns of the critical current as functions of the total magnetic flux through the junction $I_c = I_c(H)$. 
 Upper panel: homogeneous junction (dashed curve) and in the presence of ferromagnetic (FM) vortex (solid curve) 
 described by the $\theta(x)$ distribution (\ref{Eq:Model}). 
 Lower panel:  influence of the spin-dependent phase difference arising near the FM vortex core on the critical current. 
 $\Delta\alpha=0$ (dashed curve) and $\Delta\alpha (x)\neq 0$ given by the Eq. (\ref{Eq:ModelAlpha}) (solid curve). 
 Spin polarization is assumed to be such that $j_+ = 0.5j_-$ for all curves. 
 The asymmetry of red solid curve $I_c(H) \neq I_c(-H)$ shows the presence of spontaneous superconducting currents. }
 \end{figure}
   
  
The Josephson effect in the setup shown in Fig.(\ref{Fig:SFSmodel})a can be analysed using the 
formalism of Eqs.(\ref{Eq:UsadelGen},\ref{Eq:ChargeCurrentDiff},\ref{Eq:BerezinskyOP}). 
Due to the small distance between superconducting electrodes
 one can use the model 1D system shown in Fig.\ref{Fig:SFSmodel}b which 
 is the long junction with gap
functions in the leads given by (\ref{Eq:BerezinskyOP}). There are two key
points to understand the experimental results. First, the
effective gap amplitude is $\Delta_\sigma \propto \sin\theta$
that is determined by the angle $\theta$ between $\bm m$ and $\bm h_s$. As shown by
the magnetization pattern data\cite{AartsExperiment}, $\theta$ changes quite
strongly along the junction both due to the vortex-like pattern in F and the distortions of the mono-domain state 
in F' near the trench that cuts S and F' layers. The model distribution of $\theta$ along the Josephson junction 
that is used in calculations is shown in Fig.\ref{Fig:SFSmodel}c and described by 
 $\theta (x) = (\pi/2) \tilde\theta(x/L_J)$, where
 \begin{equation} \label{Eq:Model}
 \tilde\theta (\tilde x) =  \tanh\left(\frac{\tilde x}{w_c}\right) 
 \tanh\left( \frac{\tilde x -0.5}{w_e}  \right) \tanh\left( \frac{\tilde x + 0.5}{w_e} \right). 
 \end{equation}
 The basic features of this profile is that $\theta=0$ at the FM vortex center $x=0$ and at the edges $x=\pm L_J/2$, 
 while reaching $\pm \pi/2$ at the middle points. The dimensionless widths $w_e$ and $w_c$ determine the shape of the Fraunhofer pattern.  
 { Second, near the FM vortex core the magnetization pattern 
 is non-coplanar which generates both the gauge field $\bm Z$ and the spin-dependent
 phase $\alpha=\alpha (x)$ distribution along the junction. Using the model magnetization distribution
 $\bm m= ( m_\perp \cos\varphi,  m_\perp \sin\varphi, m_z)$, where $\varphi$ is the real-space polar angle
 one obtains $\bm Z = m_z \nabla\varphi /2$
 and $\alpha = \arctan [m_z \tan (\varphi - \varphi_0)]$, where $\varphi_0$ is the azimuthal angle of 
 $\bm h_s = h_s (\cos\varphi_0, \sin\varphi_0, 0)$. Note that for $|m_z|=1$ the superfluid velocity (\ref{Eq:SuperfluidVelocity})
 is zero $\bm V_s = 0$. However this is not the case at $|m_z |< 1$. The effective phase difference can be found integrating 
 the superfluid velocity across the junction $\Delta\alpha = \int_{-d}^{d} dy V_{Sy} $, where $2d$ is the junction width. 
 For $\varphi_0=0$ and provided that $m_z\approx const$ at the scale $|y|<d$  one obtains in this way
 \begin{equation} \label{Eq:ModelAlpha}
 \Delta\alpha (x)= 2[ \arctan(m_z \tan\varphi) - m_z \varphi ]
 \end{equation}
 where $\varphi = \arctan(d/x)$. 
    
   }

The Josephson current can be calculated analytically assuming that the
proximity effect is weak and one can use the linearized theory formulated for the component 
 $f_\sigma = {\rm Tr}[(\tau_x + i\tau_y)\hat g_\sigma]/2$. 
The linearized Usadel equation and expression for the current across the junction read
   \begin{align} \label{Eq:UsadelLinear}
   & D_\sigma \hat\nabla^2 f_\sigma - 
   2 \omega f_\sigma - \Delta_\sigma =0 
   \\ \label{Eq:CurrentLinear}
   & \bm j = \pi T e \sum_{\sigma=\pm } \sum_\omega 
   \nu_\sigma D_\sigma {\rm Im} (f_\sigma \hat\nabla f_\sigma^*)
   \end{align}
   
 Since the distance between superconducting electrodes is small,
 proximity-induced vortices\cite{Cuevas2007,Alidoust2013,Alidoust2015,Alidoust2017} cannot form in the junction. 
 Then the order parameter distribution can be approximated by the step-wise function
 \begin{equation} \label{Eq:DeltaModel}
 \Delta_\sigma(y) =  h_\perp \Delta_{F\sigma}
 [\cos(\chi_\sigma/2) + i Step(y) \sin(\chi_\sigma/2)].
 \end{equation}
 It is convenient to choose the gauge so that $A_y= 0$ and neglect $A_x$ component due to the small
 junction width. Then the total 
 phase difference is given by 
 $\chi_\sigma (x) =\Delta \chi + \sigma\Delta \alpha (x) + \phi x/L_J$, where  $\Delta \chi= \chi_R - \chi_L$ and
 $\Delta \alpha (x)= \alpha_R - \alpha_L$ are the usual and spin-dependent phase differences, 
 $\phi = \Phi/\Phi_0$ where $\Phi$ is the total magnetic flux through the junction area including 
 the leads, $\Phi_0$ is the flux quantum. 
The field $\bm Z$ is treated as locally homogeneous and gauged it out by using the effective phase difference (\ref{Eq:ModelAlpha}).

Within the above assumptions the linearized Usadel Eq.(\ref{Eq:UsadelLinear}) can be solved 
analytically and Eq.(\ref{Eq:CurrentLinear}) yields the following  expression for the Josephson current density
 \begin{align} \label{Eq:CurrentDensity}
 &j (x) = \sum_{\sigma= \pm } j_\sigma (\sin\theta)^2 \sin\chi_\sigma ,
 \\
 &j_\sigma = \frac{\pi T e \nu_\sigma D_\sigma}{8} \sum_\omega 
 \frac{\Delta_{F\sigma}^2}{\xi_{N\sigma}\omega^2 } , 
 \end{align}
 where
 $\xi_{N\sigma} = \sqrt{2|\omega|/D_\sigma}$. The total critical current is the given by $I_c = \sqrt{I_1^2 + I_2^2}$, where 
 \begin{align} \label{Eq:I1}
 I_1 = \sum_\sigma j_\sigma \int_{-L_J/2}^{L_J/2} (\sin\theta)^2 
 \cos(\phi x + \sigma\Delta\alpha )dx 
 \\ \label{Eq:I2}
 I_2 = \sum_\sigma j_\sigma \int_{-L_J/2}^{L_J/2} (\sin\theta)^2 
 \sin(\phi x + \sigma\Delta\alpha )dx .
 \end{align}  
 
  Fig.(\ref{Fig:Jos}) presents the dependencies $I_c = I_c (\phi)$ calculated according to the Eqs.(\ref{Eq:I1},\ref{Eq:I2}) with the help model distributions 
  (\ref{Eq:Model},\ref{Eq:ModelAlpha}) for $\theta(x)$ and $\Delta\alpha(x)$, respectively. 
The upper panel shows the standard single-slit pattern from the homogeneous junction (dashed curve) and the double-slit pattern produced by the 
junction with FM vortex (solid curve). Here the spin-dependent phase is absent $\Delta\alpha=0$.  
The distribution of $\theta (x) $ is taken in the form (\ref{Eq:Model}) with free parameters $w_{e,c}$. 
The homogeneous case is obtained in the limit $w_e, w_c \to 0$. Increasing the widths $w_{e,v}$ leads to the gradual transform of the interference pattern which at $w_e=w_c= 0.3$ acquires the double-slit form shown in the upper panel of the  Fig.(\ref{Fig:Jos}).
 Note that it is necessary to take into account the suppression of 
$\theta(x)$ both at the FM vortex core and at the boundary. In case when $w_e\to 0$ the interference pattern tends to the deformed homogeneous picture.

 The spin-dependent phase  $\Delta\alpha (x)$ can produce the   
 Josephson current (\ref{Eq:CurrentDensity}) even in the absence of the usual phase difference and 
 external magnetic field. Since the phase shift has opposite signs in spin-up/down subbands the net effect on the current shows up only due to the 
 certain amount of spin-filtering in the system\cite{Silaev2017,BobkovaSilaev2017}, that is
 $j_+ \neq j_-$. This is possible only if the spin-up and spin-down diffusion coefficients and/or 
 DOS are different. The model distribution (\ref{Eq:ModelAlpha}) produced by the FM vortex leads to the non-trivial modification of
 the Franhofer pattern, shown by the solid line in Fig.(\ref{Fig:Jos}), lower panel. The parameters here are 
 $d= 0.2 L_J$, $j_- = 0.5 j_+$. Asymmetric curve $I_c(H)\neq I_c(-H)$ is qualitatively similar to the one produced by the 
 localized magnetic field or the internal  phase shifts  in junctions between chiral $p$-wave superconductors\cite{Kidwingira2006}.
 Here the asymmetry is finite due to the spin filtering $j_+ \neq j_-$ and thus it signal the presence of spontaneous currents.

 {  For the the S/F'/F setup with magnetic configurations similar to Fig.(\ref{Fig:SFSmodel})a the effect of spontaneous current is rather tiny
 since the induced superfluid velocity is zero $\bm V_s = 0$ both at the center where  $m_z = 0$ and outside the FM vortex core 
 where $m_z=1$. 
 The situation becomes completely different for $ \bm z \cdot \bm h_s \neq 0$. For example if 
 $\bm h_s = h_s \bm z$ one can see that the spin-dependent phase is constant $\nabla \alpha =0$ so that 
 $\bm V_s = \sigma m_z \nabla \varphi /2 $ and has singularity at $r=0$.  The current is however 
 finite $j \sim h_\perp^2/r$ since $h_\perp = \sqrt{1-m_z^2} \sim r$. This behaviour is similar to the 
 orbital supercurrents around Abrikosov vortices in bulk superconductors.  Note that this texture-induced spontaneous 
 current is not related to the anomalous Josephson effect \cite{Braude2007,Buzdin2008, Reynoso2008,Grein2009, Zazunov2009, Liu2010, Mironov2015, Brunetti2013, 
Yokoyama2014, Kulagina2014,Nesterov2016, Konschelle2015,
Moor2015, Moor2015a,Bobkova2016, Szombati2016, Silaev2017}   since it exists without the weak link in the superconducting layer. 
 The singularity of $\bm Z$ can be removed by the gauge transform introducing the vorticity to
 the spin-dependent phase $\alpha = \varphi$. Therefore in such system FM vortex 
 generates singly-quantized superconducting vortices in the proximity-induced Berezinskii superconductor.
 The physics of such vortices and superconducting kinks that are generated by magnetic domain walls in the 
 multilayer setups similar to the Fig.(\ref{Fig:SFSmodel})a  is potentially quite rich  
 but is beyond the scope of the present paper. 
  }

  To summarize, in this letter I explain the recent observation of the unusual magnetic field dependence of
   critical current of the Josephson junction through ferromagnetic vortex. 
  The key theoretical finding is that the experimental results are consistent with the existence of 
  proximity-induced Berezinskii superconductors in the S/F'/F systems with non-collinear strong ferromagnets. 
  The effective order parameter amplitude is defined by the angle between magnetic moments in F and F' layers.
  The inhomogeneous distribution of this angle results in the double-slit interference  
  pattern of the critical current. 
  Besides that, the non-coplanar magnetic texture near the FM vortex core generates 
  spin-dependent phase gradients and the emergent gauge field which can be combined into the invariant 
  combination yielding the spin-dependent part of the superfluid velocity $\bm V_s$.  
  The lifted degeneracy of DOS and diffusion coefficients in spin subbands in strong ferromagnets
   converts $\bm V_s$ to the spontaneous supercurrent  which is shown to result in the critical current asymmetry $I(H) \neq I(-H) $. 
  
 Finally, I show that FM vortex can induce spontaneous vorticity in the Berezinskii state, 
 provided that the magnetization in F' layer has an out-of-plane direction. That finding demonstrates that 
 besides affecting transport properties of textured magnets\cite{Volovik1987, Nagaosa2013} the emergent gauge field can show up in the 
 orbital motion of spin-triplet Cooper pairs and even produce topological defects in the superconducting order parameter .
 The result obtained here for the FM vortex are in general valid for magnetic skyrmions as well, owning to the similarity in their
 magnetization distributions.   
    
 I thank T. T Heikkil\"a ,  A. Mel'nikov, I. Bobkova and A. Bobkov
 for stimulating discussions.
  The work was supported by the Academy of Finland.

%


\end{document}